\documentclass[a4paper,article,twocolumn,english,twocolumns,epl]{revtex4}
\usepackage{amsmath}
\usepackage{amsfonts}
\usepackage{amssymb}
\usepackage{comment}
\usepackage{color}
\usepackage{graphicx}
\usepackage{color}
\definecolor{darkblue}{rgb}{0,0,0.6}
\definecolor{darkred}{rgb}{0.6,0,0}
\definecolor{darkgrey}{rgb}{0.6,0.6,0.6}
\usepackage[colorlinks=true,urlcolor=darkblue,citecolor=darkblue,linkcolor=darkred,hyperfootnotes=false]{hyperref}

\providecommand{\U}[1]{\protect\rule{.1in}{.1in}}
\begin{document}
\title{Instantons for the destabilization of the inner Solar System}
\author{Eric Woillez$^{1,2}$ and Freddy Bouchet$^{1}$}
\affiliation{$^1$Univ Lyon, Ens de Lyon, Univ Claude Bernard, CNRS, Laboratoire
de Physique, F-69342 Lyon, France.}
\affiliation{$^2$Department of Physics, Technion, Haifa 32000, Israel}

\begin{abstract}
For rare events, path probabilities often concentrate close to a predictable path, called instanton. First developed in statistical physics and field theory, instantons are action minimizers in a path integral representation. For chaotic deterministic systems, where no such action is known, shall we expect path probabilities to concentrate close to an instanton? We address this question for the dynamics of the terrestrial bodies of the Solar System. It is known that the destabilization of the inner Solar System might occur with a low probability, within a few hundred million years, or billion years, through a resonance between the motions of Mercury and Jupiter perihelia. In a simple deterministic model of Mercury dynamics, we show that the first exit time of such a resonance can be computed. We predict the related instanton and demonstrate that path probabilities actually concentrate close to this instanton, for events which occur within a few hundred million years. We discuss the possible implications for the actual Solar System.
\end{abstract}

\maketitle

Rare events can be very important if their large impact compensate for their low probability. From a dynamical perspective, when conditioned on the occurence of a rare event, path probabilities often concentrate close to a predictable path, called instanton. This is a key and fascinating property for the dynamics of rare events and of their impact \cite{ragone2018computation}, which was first observed in statistical physics, for the nucleation of a classical supersaturated vapor \cite{langer_1967_condensation_point}. Soon after, a similar concentration of path probabilities has been studied in gauge field theories  \cite{Coleman:1978ae,zinn1996quantum}, for instance for the Yang-Mill theory. 
Instantons continue to have number of applications in modern statistical physics, for instance to describe excitation chains at the glass transition \cite{langer2006excitation}, reaction paths in chemistry \cite{kampen_stochastic_2007}, escape of brownian particles in soft matter \cite{woillez2019escape}, MHD \cite{Berhanu_etc_Fauve_2007_EPL_MagneticFieldReversal} and turbulence \cite{grafke2013instanton,LAURIE:2015:A,grafke2015efficient,bouchet2019rare,dematteis2018rogue}, among many other examples. Moreover, a large effort has been pursued to develop dedicated numerical approches to compute instantons \cite{grafke2019numerical}. Inspired by the earlier works, action minimization have found a rigorous mathematical treatment, through the Freidlin-Wentzell large deviation theory \cite{FW2012} of ordinary differential equations with small noises \cite{Graham1987macroscopic}. 

\begin{figure}
\begin{centering}
\includegraphics[height=4cm]{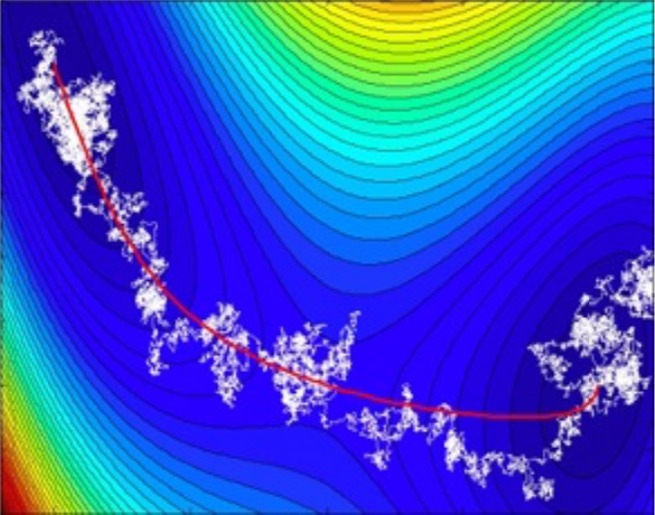}
\par\end{centering}
\caption{Instanton for a Brownian particle in a bistable potential. 
The particle's trajectory from one attractor to another (white line) closely follows the minimum action path
(instanton, red line), up to thermal fluctuations. The level curves of the potential are displayed in the
background, with the color scale giving the potential's height (courtesy Eric Vanden-Eijnden). \label{fig:instanton_example}}
\end{figure}

In all those classical or quantum applications, instantons appear as action minimizers, for a saddle point evaluation of a path integral. The basic property of the instanton phenomenology is that, conditioned on the occurence of a rare event, path probabilities concentrate close to a predictable path. Fig. (\ref{fig:instanton_example})
gives an illustration of this property for a particle in
a bistable potential.  Shall we expect this phenomenology to be valid for systems for which the Freidlin-Wentzell action (Please note that the word "action" refers to the path integral of large deviation theory, and has nothing to do with the classical action of analytical mechanics that can be written for Hamiltonian dynamics). does not exist in the first place, for instance chaotic deterministic systems? The main aim of this work is to open this fascinating question for a paradigmatic problem in the history of physics: the dynamics of the Solar System. Shall we expect an instanton phenomenology for rare events that shaped or will shape the Solar System history?

The discovery that our solar system is chaotic with a Lyapunov time
of about $5$ million years \cite{laskar1989numerical,laskar1990chaotic,Sussman_Wisdom_1992_chaotic}
has disproved the previous belief that planetary motion would be predictable
with any desired degree of precision. On the contrary, chaotic motion
sets an horizon of predictability of a few tens of million years for
the solar system. Even more striking has been the discovery that about 1\%
of the trajectories in the Solar System lead to collisions between planets, or between
planets and the Sun within $5$ billion years \cite{laskar2009existence}.
As shown numerically, chaotic disintegration of the inner solar system
(i.e. the four terrestrial planets) always happens through a resonance between
the motion of Mercury's and Jupiter's perihelia \cite{batygin2008dynamical,laskar2008chaotic,laskar2009existence,boue2012simple}, related to a large increase in Mercury's eccentricity. Stochastic perturbation to planetary motion exists, for instance through the chaotic motion of the asteroid belt, but is too weak to be responsible
for the rare destabilizations of the inner solar system \cite{laskar2008chaotic,woillez2017long}. Instead, stochasticity in the solar system appears because of the development of internal deterministic chaos \cite{laskar2008chaotic}.

Does an instanton phenomenology exist for the rare destabilization of the Solar System? Our first result will be obtained within a simplified model of Mercury's dynamics \cite{batygin2015chaotic}. We predict for this model the probability distribution of the first destabilization time, the instanton paths, and check the instanton phenomenology.

The secular dynamics describes the planetary motion averaged over fast orbital motion. The secular dynamics Hamiltonian is
\begin{equation}
H(\mathbf{I},\mathbf{\Phi})=H_{int}(\mathbf{I})+\underset{\mathbf{k}\in\mathbb{Z}^{16}}{\sum}A^{k}(\mathbf{I})\cos\left(\mathbf{k}.\mathbf{\Phi}\right),\label{eq:complete secular H}
\end{equation}
where $(\mathbf{I},\mathbf{\Phi})$ is the canonical set of Poincar\'e action-angle
variables for the 8 planets, $\mathbf{k}$ is a vector of integers,
and the coefficients $A^{k}$ are functions of the action variables
only (see e.g \cite{laskar1995stability} for the explicit expression of $H$ to forth order in planetary eccentricities and inclinations). We will study Mercury's possible destabilization in the framework of a simplified model proposed by Batygin and col. \cite{batygin2015chaotic}. This model should be seen as a minimal model retaining the relevant interactions leading to destabilization of the inner Solar System but is not expected to describe quantitatively the inner Solar System. 

The approximations of \cite{batygin2015chaotic} consist in keeping only the
degrees of freedom of a massless Mercury in the Hamiltonian (\ref{eq:complete secular H}),
and replace all other action-angle variables by their quasiperiodic
approximation. Assuming moreover that only a small number of periodic
terms in Eq. (\ref{eq:complete secular H}) significantly affect the
long-term secular motion of Mercury \cite{batygin2008dynamical,lithwick2011theory,boue2012simple,batygin2015chaotic},
Mercury's simplified Hamiltonian is 
\begin{eqnarray}
H & = & H_{int}(I,J)+E_2\sqrt{I}\cos\left(\varphi\right)+S_2\sqrt{J}\cos\left(\psi\right)\nonumber \\
 &  & +E_T\sqrt{I}\cos\left(\varphi+\left(g_{2}-g_{5}\right)t+\beta\right),\label{eq:Hamiltonien BMH}
\end{eqnarray}
where $\varphi$ and $\psi$ are the canonical angles conjugated
to $I=1-\sqrt{1-e^{2}}$ and $J=\sqrt{1-e^{2}}\left(1-\cos i\right)$ respectively,
and $e$ and $i$ are Mercury's eccentricity and inclination \cite{batygin2015chaotic}.
$g_{5}$, $g_{2}$ and $s_{2}$ are frequencies involved
in the quasiperiodic decomposition of the motion of Jupiter ($g_{5}$) and Venus ($g_{2}$ and $s_{2}$).
The numerical values for the other coefficients in Eq. (\ref{eq:Hamiltonien BMH})
are given in~appendix. 

\noindent\emph{A slow variable for Mercury's dynamics:} We first show how
a slow variable can be built from the dynamics defined by the
Hamiltonian (\ref{eq:Hamiltonien BMH}). In Eq. (\ref{eq:Hamiltonien BMH}),
 $H_{int}$ only depends on the actions. Would the total Hamiltonian
be reduced to this part, the
actions would be constant and the canonical angles would simply grow
linearly with time according to Hamilton's equations
\begin{equation}
\begin{cases}
\dot{\varphi}(t) & =\frac{\partial H_{int}}{\partial I}=-g_{1}(I,J)+g_{5},\label{eq:integrable motion}\\
\dot{\psi}(t) & =\frac{\partial H_{int}}{\partial J}=-s_{1}(I,J)+s_{2}.
\end{cases}
\end{equation}
The  fundamental frequencies $g_{1}\left(I,J\right)$ and $s_{1}\left(I,J\right)$ describe Mercury's perihelion
precession at frequency $g_{1}$, and its orbital plane oscillations with respect to the invariant reference plane,
at frequency $s_{1}$. For the model (\ref{eq:Hamiltonien BMH}),  $g_{1}$ value is about $5.7''/yr$, corresponding
to a period of about $227000$ years (This value is actually specific of our model. The current value of $g_1$ for the real Solar System would be about $5.60"/yr$).

\begin{figure}
\begin{centering}
\includegraphics[height=5cm]{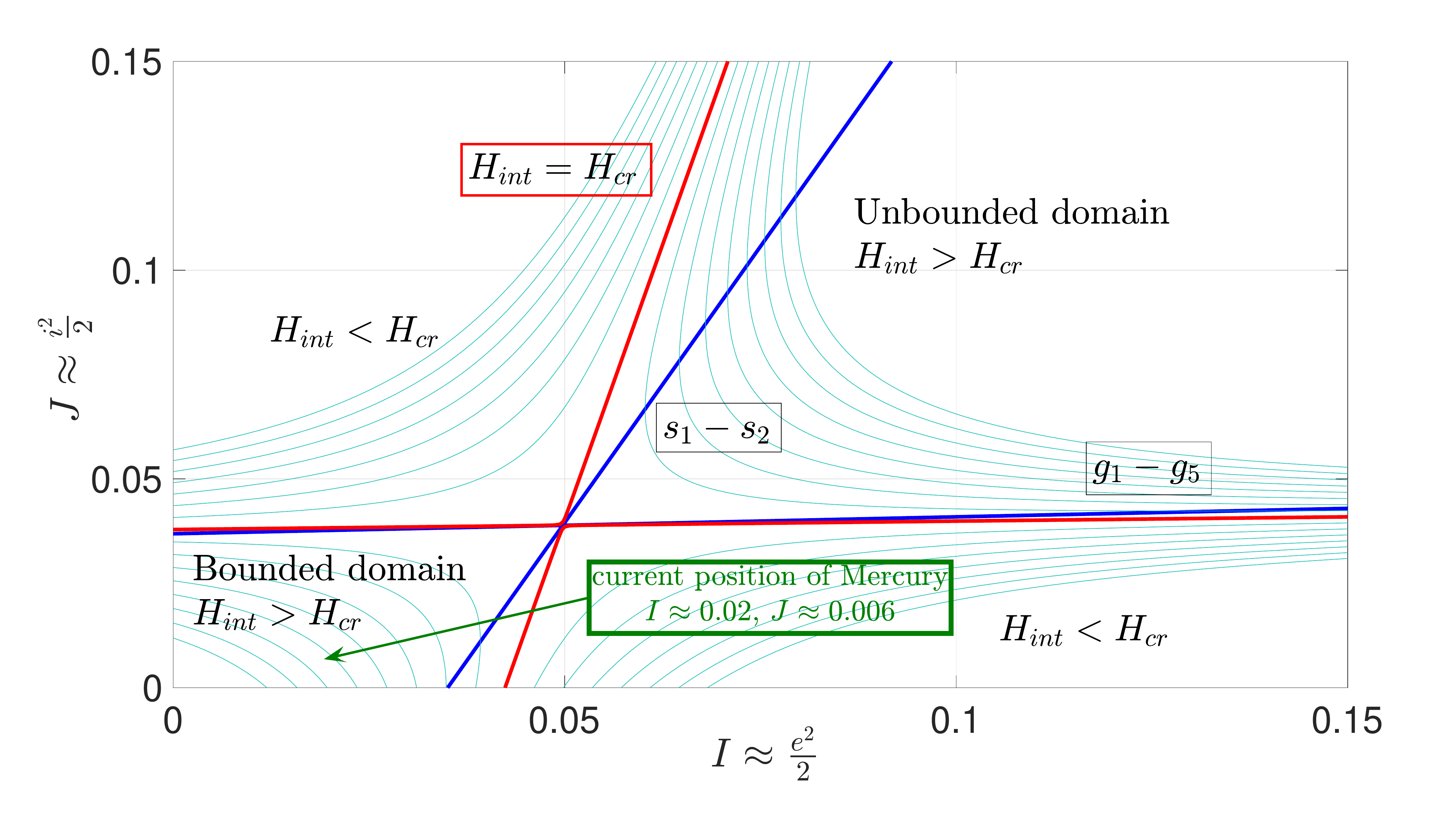}
\par\end{centering}
\caption{Level curves of $H_{int}(I,J)$ in action space. The surface defined by
$H_{int}(I,J)$ has the structure of a saddle. Mercury currently satisfies $H_{int}>H_{cr}$ and is located in the bounded domain. For destabilization to occur,
Mercury has to cross the saddle and enter the unbounded domain. \label{fig: level H}}
\end{figure}

Through the chaotic dynamics of (\ref{eq:Hamiltonien BMH}), the fundamental frequencies $\left\{g_{1},s_{1}\right\}$ change over time. Mercury's secular motion might enter into resonance with the external periodic forcing if $g_{1}$ or $s_{1}$ comes
close to one of the frequencies $g_{5}$, $g_{2}$ or $s_{2}$. In
particular, the Mercury-Jupiter perihelion resonance, between $g_{1}$ and $g_{5}$, might trigger Mercury's destabilization \cite{batygin2008dynamical,laskar2008chaotic,laskar2009existence,boue2012simple}.
The three curves of equations $g_{1}\left(I,J\right)=g_{5}$, $s_{1}\left(I,J\right)=s_{2}$
and $g_{1}\left(I,J\right)=g_{2}$ can be represented in the $(I,J)$
plane, together with the current values of Mercury's action variables.
We obtain in Fig. (\ref{fig: level H}) the so-called "resonance
map" which is now widely used for weakly non-integrable
systems \cite{laskar1993frequency,morbidelli2002modern}. 
We write (\ref{eq:Hamiltonien BMH}) as $H=\widetilde{H}+H_{pert}$,
with
\begin{eqnarray}
&\widetilde{H}=H_{int}+E_2\sqrt{I}\cos\left(\varphi\right)+S_2\sqrt{J}\cos\left(\psi\right), \label{eq:def slow hamiltonian-1}\\
&H_{pert}=E_T\sqrt{I}\cos\left(\varphi+\left(g_{2}-g_{5}\right)t+\beta\right). \label{eq:Hpert}
\end{eqnarray}
The term $H_{pert}$ given by (\ref{eq:Hpert})
creates a weak perturbation for Mercury's long-term evolution. To
find the order of magnitude at which $H_{pert}$ affects the long-term
dynamics of Mercury, we employ Lie transform methods \cite{morbidelli2002modern} with the software TRIP (TRIP is a general computer algebra system dedicated to celestial mechanics developed at the IMCCE (Copyright 1988-2019, J. Laskar ASD/IMCCE/CNRS). TRIP is particularly efficient to handle series with a large number of terms like those usually appearing in Lie transforms. ).

There exists new
action-angle variables and a canonical transformation such that Mercury's
Hamiltonian can be put in the form
\begin{equation}
H'=\widetilde{H}'\left(I',J',\varphi',\psi'\right)+H_{pert}'\left(I',J',\varphi',\psi',\left(g_{2}-g_{5}\right)t\right),\label{eq:hamiltonian transform}
\end{equation}
where the order of magnitude of $H_{pert}'$ is much smaller
than $H_{pert}$. The Lie transform creates periodic terms in $H_{pert}'$ that contain new combinations of the angles $\varphi',\psi'$ and $\left(g_{2}-g_{5}\right)t$  (given in the appendix). The difference between $H_{pert}$ and $H_{pert}'$ is that the angular terms of the latter are resonant, which means that their frequencies can vanish. The existence of such resonant terms, even of small amplitude, generate long-term chaotic motion.

The Hamiltonian
(\ref{eq:hamiltonian transform}) defines a dynamical system with two well separated time scales. On a time scale of the order of $\frac{1}{g_1}$, the action-angle variables evolve according to Hamilton's equations of motion. The flow is chaotic with a Lyapunov time $\tau_{L}$ of the order of one million years \cite{batygin2015chaotic}. 
$\widetilde{H}'$ evolution
\begin{equation}
\dot{\widetilde{H}'}=\left\{ H_{pert}',\widetilde{H}'\right\} ,\label{eq:motion good slow}
\end{equation}
sets a new time scale. In Eq. (\ref{eq:motion good slow}), the notation $\left\{ \right\} $ represents the canonical Poisson brackets. Eq. (\ref{eq:motion good slow}). shows that $\widetilde{H}'$ is a slow variable, because its time evolution is driven by  $H_{pert}'\ll H_{pert}$. As will become clear in the following, $\widetilde{H}'$ remains almost  constant on the fast time scale, and has only significant variations on a timescale of a few hundred million years.

\noindent\emph{Diffusion of the slow variable:} The theory of white noise limit
for slow-fast dynamical systems (see e.g. \cite{gardiner1985stochastic})
suggests that on a timescale much larger than $\tau_{L}$, Eq.
(\ref{eq:motion good slow}) is equivalent to a diffusion process. This limit is valid assuming that the variations of $\widetilde{H}'$ on the timescale $\tau_{L}$ are sufficiently small. Two additional phenomenological approximations can be made: first, numerical simulations performed with Eq. (\ref{eq:motion good slow})
show that the drift is very small compared to the diffusion coefficient,
and can be neglected. Second, the range of $\widetilde{H}'$ values before destabilization is small, and the diffusion coefficient can be considered as constant. The long-term evolution of 
$\widetilde{H}'$ can thus be modeled by the standard Brownian motion
\begin{equation}
\dot{\widetilde{H}'}=\sqrt{D}\xi(t),\label{eq:diffusion slow}
\end{equation}
where $\xi(t)$ is the Gaussian white noise  with correlation
function $\left\langle \xi(t)\xi(t')\right\rangle=\delta(t-t') $.
Unfortunately, the exact expression for $D$
involves the full correlation function of the Hamiltonian flow defined by $\widetilde{H}'$. It is too intricate to be useful in practice. Starting from the formal expression, it is shown in the appendix that an order
of magnitude is
\begin{equation}
D\approx 2\left|H_{pert}\right|^{6}\tau_{L}/|\widetilde{H}|^{4},\label{eq:simple exp for D}
\end{equation}
where $\left|\widetilde{H}\right|$ and $\left|H_{pert}\right|$ are orders of magnitude of (\ref{eq:Hpert}) and (\ref{eq:def slow hamiltonian-1})
respectively. Eq. (\ref{eq:simple exp for D}) is our first important result. Evaluating Eq. (\ref{eq:simple exp for D}) gives $D\approx7.2\times10^{-7}\;Myr^{-3}$. The associated diffusion time scale for $\widetilde{H}'$ is evaluated to one billion years. Those results justifies the self-consistency of the choice for the slow variable.

\noindent\emph{Distribution of the first destabilization times of Mercury: } We now discuss qualitatively the implications of the existence of a slow variable for Mercury's destabilization. This discussion is best understood looking at the level curves of $H_{int}(I,J)$
in action space displayed in Fig. (\ref{fig: level H}). It can be seen that the landscape defined by
$H_{int}(I,J)$ has the topology of a saddle. The saddle is exactly located at the intersection between the two resonances $g_{1}-g_{5}$
and $s_{1}-s_{2}$, with the value $H_{int}=H_{cr}$. The domain of equation
$H_{int}(I,J)\geq H_{cr}$ has two disjoint components, one bounded (bottom left) and the other unbounded (top right), only
connected by the saddle point $(I_{cr},J_{cr})$. The initial orbital parameters
of Mercury $e$ and $i$ are located in the bounded domain, which implies that the short-time orbital fluctuations are restricted to this part of phase space. When  $H_{int}$ reaches the value $H_{cr}$, Mercury can cross the saddle and enter the unbounded domain of phase space. This latter event defines Mercury's destabilization.

We explain in the appendix how the above simple criterion translates into an equivalent criterion for $\widetilde{H}'$: there exists a threshold $h_{cr}$ for which the first destabilization time exactly corresponds to the first hitting time of $\widetilde{H}'$ to $h_{cr}$.

\begin{figure}
\begin{centering}
\includegraphics[height=4cm]{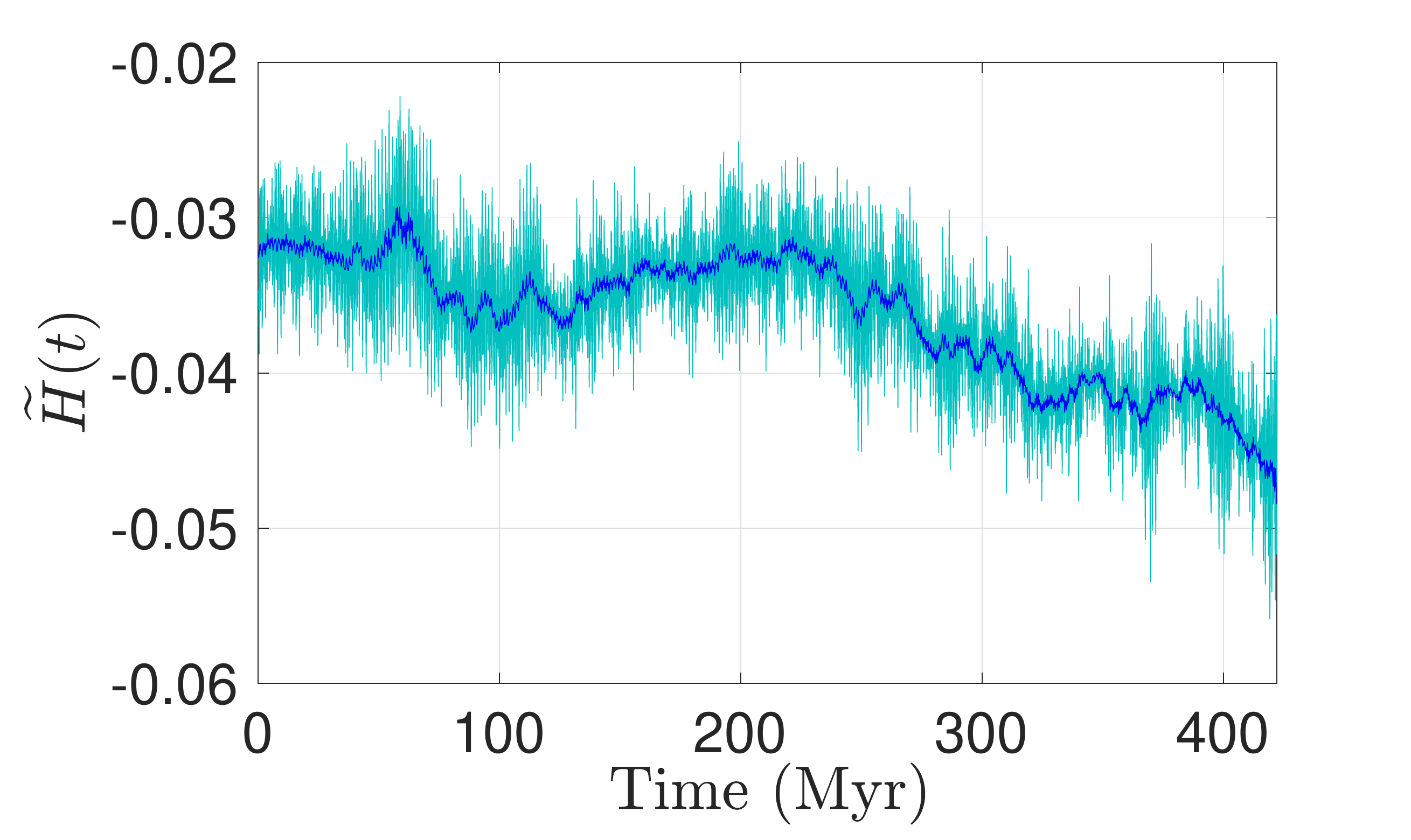}
\par\end{centering}
\caption{A trajectory $\widetilde{H}(t)$ (cyan) compared to its local time
average $h(t)$ (blue). The local time averaging of $\widetilde{H}(t)$
suppresses the fast oscillations that do not correspond to long-term
variations. For the long-term chaotic dynamics, $h(t)$ is an slow variable that follows a standard Brownian motion. \label{fig:Htild versus average}}
\end{figure}
The full expression of $\widetilde{H}'$ is  an intricate serie composed of a large number of periodic terms of small amplitude, which explicit expression is difficult to handle. Following \cite{batygin2015chaotic}, we prefer to use  in practice the local time average $h(t)=\left\langle \widetilde{H}\right\rangle _{\left[t-\theta,t+\theta\right]}$ as an approximation of $\widetilde{H}'$, which is much simpler to implement numerically.
The time frame $\theta$ has to be much larger than the frequency of the fast variations of $\widetilde{H}$ given by the frequency $g_{2}-g_{5}$ according to Eq. (\ref{eq:Hpert}).  As an example, the time variations of $\widetilde{H}(t)$
compared to those of $h(t)$ is displayed in Fig. (\ref{fig:Htild versus average}) with $\theta=2\; Myr$. We then identify the
diffusion Eq. (\ref{eq:diffusion slow}) for $\widetilde{H}'$
and that for $h$.

Tracking numerically the value of $h(t)$ of trajectories leading to destabilization confirms that the distribution $h(\tau)$ (where $\tau$ is the destabilization time) is peaked at the value $h_{cr}=-0.048$, which can thus be identified as the destabilization threshold. We must also add
a reflective boundary for a upper value $h_{sup}$, accounting
for the fact that the chaotic region of phase space before destabilization is bounded. Destabilization
of Mercury occurs when the Brownian motion defined by $h(t)$ reaches $h_{cr}$. For a standard Brownian motion, the distribution $\rho(\tau)$ of first hitting times of the value $h_{cr}$ can be derived exactly (see appendix). 
\begin{figure}
\begin{centering}
\includegraphics[width=6cm]{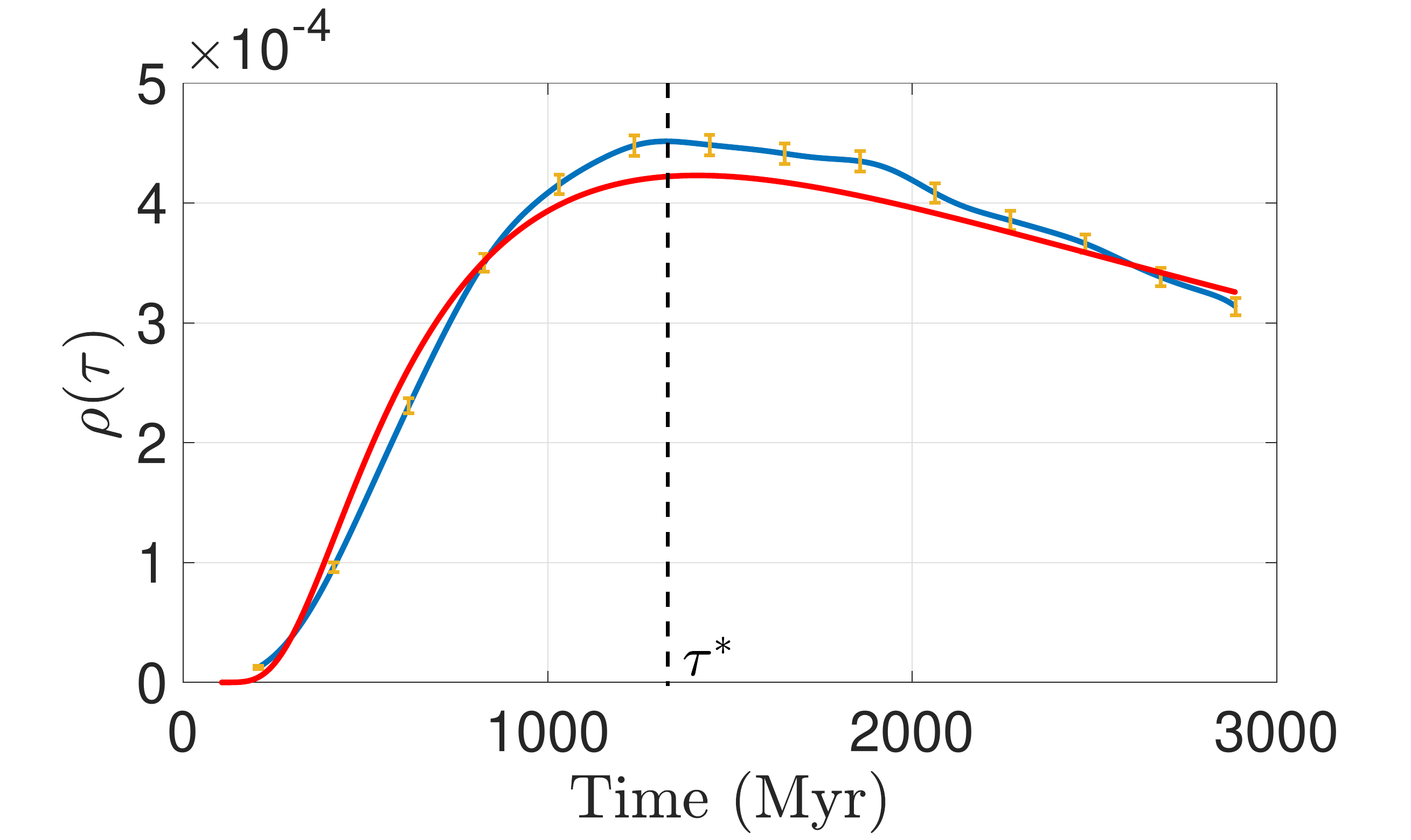}
\par\end{centering}
\caption{Probability distribution of Mercury's first destabilization time.
The distribution
of Mercury's first destabilization time is computed with a direct
numerical simulation (blue curve) and with the theoretical prediction
of the diffusive model Eq. (\ref{eq:diffusion slow}) (red curve). \label{fig:PDF_sortie}}
\end{figure}
The latter
is displayed in Fig. (\ref{fig:PDF_sortie}), together with the distribution
obtained from direct numerical simulations of Hamilton's equations. $D$ is the only fitting parameter and can be estimated as
$
D\approx9.6*10^{-7}\:Myr^{-3}
$.
Using this value, Fig. (\ref{fig:PDF_sortie}) shows that
the diffusive model Eq. (\ref{eq:diffusion slow}) gives a excellent qualitative agreement with the direct
numerical simulations. The fitted value of $D$ is also in agreement with Eq. (\ref{eq:simple exp for D}) and its order of magnitude $D\approx7.2\times10^{-7}\;Myr^{-3}$. 

\noindent\emph{Instanton paths for Mercury:} We now focus on the probability
that Mercury's orbit is destabilized in short times $\tau_{L}\ll\tau\ll\tau^{*}$, where $\tau^{*}$ is the maximum of $\rho(\tau)$. 
The probability $\mathbb{P}(\tau)=\int_{0}^{\tau}\rho_{th}(\tau'){\rm d}\tau'$
that the destabilization of Mercury's orbit occurs in a time shorter than $\tau$ 
is dominated at short times by the exponential term
$
\rho(\tau)\underset{\tau\rightarrow0}{\asymp}e^{-\frac{\bar{\tau}}{\tau}},\label{eq:ldp short times}
$
where $\bar{\tau}=\frac{\left(h_{0}-h_{cr}\right)^{2}}{4D}\approx1.56*10^{9}$
years.
%

\begin{figure}
\includegraphics[width=7cm]{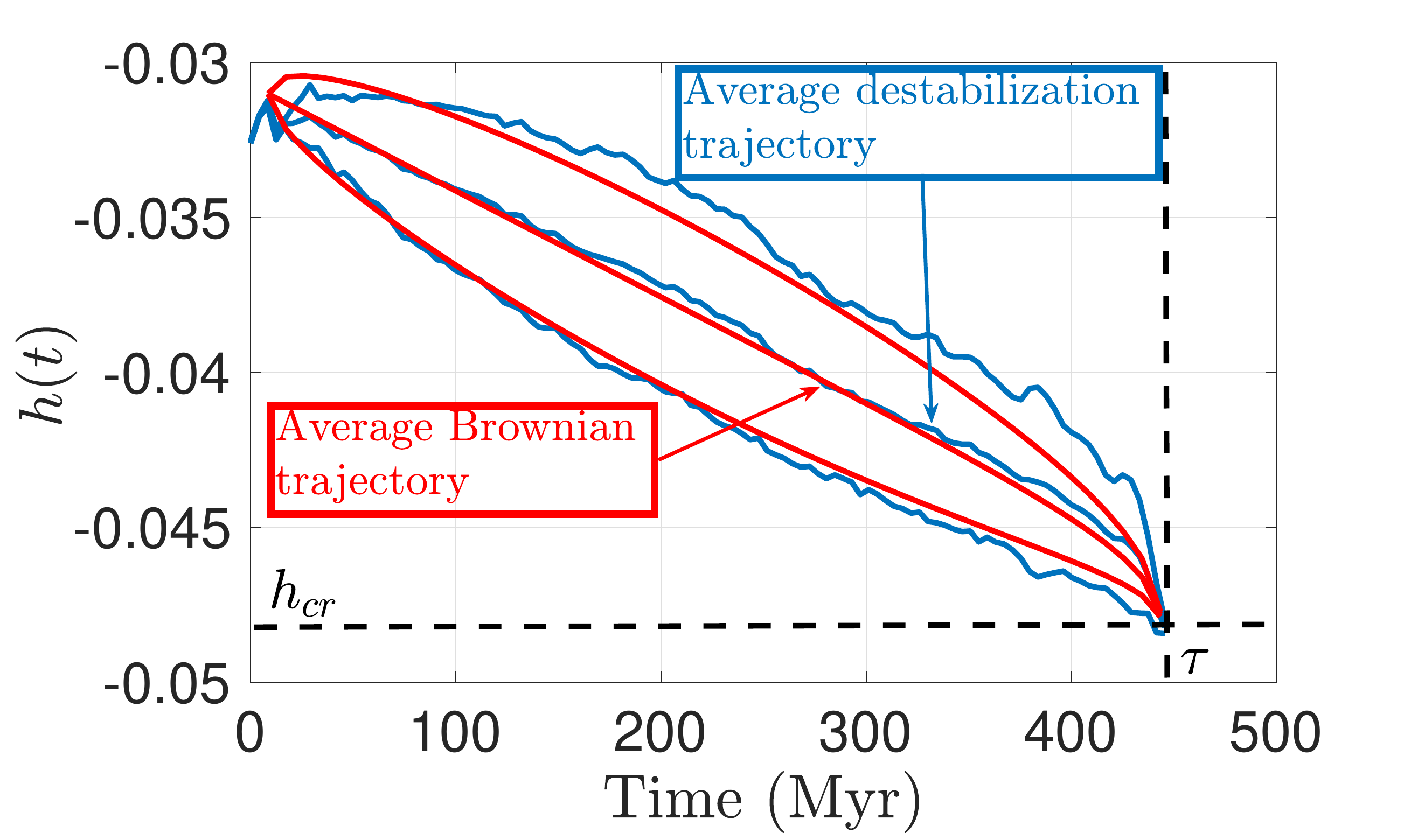}
\caption{Prediction of the trajectory leading to Mercury's short-term destabilization.
The blue curves display the average trajectory and the variance of
the trajectories leading to a destabilization at $\tau=445$ million years, obtained
with direct numerical simulations. The red curves display the same
quantities obtained with the theory of rare events (prediction of
the instanton, see appendix). \label{fig:instantons}}
\end{figure}

The exponential growth is the signature that short-term destabilizations
of Mercury are rare events. The slow variable $h(t)$, conditioned
on the fact that destabilization occurs at a given time $\tau$, is
predictable by the instanton path. The dynamics of $h(t)$ is simple enough such that the instanton path
can be computed exactly: it is the straight path starting at
$h(0)$ and reaching $h_{cr}$ at time $\tau$. We can even obtain a more precise result, namely the exact expressions for the average and the variance
of all trajectories destabilized in a given time $\tau$. The theoretical
and numerical results for $\tau=445$ million years
is displayed in Fig. (\ref{fig:instantons}). The middle blue curve displays the averaged trajectory obtained through direct numerical averaging of all trajectories leading to destabilization at time $\tau$. In addition, the upper and lower blue curves display the variance of the trajectories ensemble, and show how the trajectories depart from the most probable trajectory. We have superimposed three red curves that represent the average and variance of the probability distribution $\bold{P}[h,t|(h_{cr},\tau),(h_0,0)]$ to observe the value $h$ at time $t$, with the constrain $h(\tau)=h_{cr}$, for the standard Brownian motion $h(t)$.

The agreement between
the diffusive model of $h$ and Mercury's dynamics  can be considered as excellent, notwithstanding the small discrepancy at short times coming from the finite correlation time of Mercury's secular dynamics. This is a second
confirmation that the diffusive model for the slow
variable is consistent both for the prediction of Mercury's first
destabilization time distribution, and for the prediction of instantons. However, we note that the simple picture of a straight-line instanton is bound to the validity of the diffusive limit used to derive Eq. (\ref{eq:diffusion slow}). The simple approach described in this paper would fail if, for example, the averaged dynamics of $\widetilde{H}'$ would not be negligible.\\

Within the Batygin--Morbidelli--Holman dynamics, a reduced model of the inner Solar System with deterministic chaos, we have shown that the first exit time for a Mercury-Jupiter resonance can be computed from an effective stochastic diffusion. We have gone  beyond this result, and we predicted the related instanton and demonstrated that path probabilities actually concentrate close to this instanton, for events which occur within a few hundred million years. For the Batygin--Morbidelli--Holman model, both the instanton and the variance of the trajectories leading to Mercury's destabilization can be computed exactly. While the model contains some of the features of the inner Solar System dynamics, it neglects others. Clearly, this model should not be expected to quantitatively predict first exit times for the actual Solar System. Nevertheless, the instanton phenomenology is robust to more complex dynamics. Even if the secular dynamics of the real Mercury cannot be reduced to a simple diffusion model as done in this paper, our striking results suggest that the destabilization of the Solar System might indeed occur though an instanton phenomenology. Our work opens this question, which should be addressed within other models, that have to be realistic enough for describing faithfully the actual dynamical mechanisms, but simple enough for a proper statistical study.

\begin{acknowledgments} We are highly indebted to F. Mogavero and J. Laskar for their constant help all along this work and in particular for having shared with us the private version of TRIP and the quasiperiodic decomposition of planetary motion. We also thank C. Batygin and A. Morbidelli for having shared their previous results with us, and for interesting discussions. The research leading to these results has received funding from the
European Research Council under the European Union\textquoteright s
seventh Framework Program (FP7/2007-2013 Grant Agreement No. 616811).
\end{acknowledgments}
\nocite{morbidelli1997role,morbidelli2002modern,gardiner1985stochastic,li2014spin}

\bibliographystyle{plain_url}
\bibliography{biblio_article_mercure}

\begin{widetext}
\section{Coefficients of Mercury's Hamiltonian \label{sec:Explicit-expression-of}}

Mercury's simplified Hamiltonian is given by Eq. (2) in the main text
\begin{align}
H= & H_{int}(I,J)\nonumber \\
 & +E_{2}\sqrt{I}\cos\left(\varphi\right)+S_{2}\sqrt{J}\cos\left(\psi\right)\nonumber \\
 & +E_{T}\sqrt{I}\cos\left(\varphi+\left(g_{2}-g_{5}\right)t+\beta\right),\label{eq:BMH}
\end{align}
with
\begin{equation}
H_{int}\left(I,J\right)=\left(E_{1}+g_{5}\right)I+E_{3}I^{2}+\left(S_{1}+s_{2}\right)J+S_{3}J^{2}+F_{ES}IJ.\label{eq:Hint}
\end{equation}
We give in table (\ref{tab:values BMH}) the numerical value for the
coefficients.

\begin{table}[h]
\begin{centering}
\begin{tabular}{|c|c|}
\hline 
$E_{1}+g_{5}$ & $-1.68964$\tabularnewline
\hline 
$E_{3}$ & $-0.905766$\tabularnewline
\hline 
$S_{1}+s_{2}$ & $-1.54396$\tabularnewline
\hline 
$S_{3}$ & $-8.55372$\tabularnewline
\hline 
$F_{ES}$ & $45.2859$\tabularnewline
\hline 
$E_{2}$ & $0.0730504$\tabularnewline
\hline 
$S_{2}$ & $0.0421457$\tabularnewline
\hline 
$E_{T}$ & $0.0643625$\tabularnewline
\hline 
$g_{2}$ & $7.4559$\tabularnewline
\hline 
$g_{5}$ & $4.2575$\tabularnewline
\hline 
$s_{2}$ & $-6.57$\tabularnewline
\hline 
$\beta$ & $169.86^{\circ}$ \tabularnewline
\hline 
\end{tabular}
\par\end{centering}
\caption{Numerical value (in arcsec/yr) of the coefficients of the Hamiltonian
(2) in the main text. The phase $\beta$ is expressed in degrees.
\label{tab:values BMH}}
\end{table}

\section{Diffusion process for the slow variable\label{subsec:Diffusion-of-the}}

The present section is quite technical. We derive the formal expression
of the diffusion coefficient $D$ in Eq. (8) using Lie transform methods,
and we explain how a good order of magnitude for $D$ can be deduced
from the result. The computation have been done with the software
TRIP developed at the IMCCE by Jacques Laskar and Mickael Gastineau
(https://www.imcce.fr/trip/), which is precisely devoted to the computation
of series in celestial mechanics.

\subsection{List of third order resonances \label{subsec:Amplitude-of-the}}

We start from the Hamiltonian (\ref{eq:BMH}) (Eq. (2) of the main
text), that we decompose in two parts 
\begin{equation}
H=\widetilde{H}\left(I,J,\varphi,\psi\right)+\epsilon H_{pert}\left(I,J,\varphi,\psi,gt\right),,\label{eq:formal eps}
\end{equation}
where $g=g_{2}-g_{5}$, with $\widetilde{H}$ and $H_{pert}$ given
by Eqs. (4-5) of the main text
\begin{equation}
\widetilde{H}=H_{int}(I,J)+E_{2}\sqrt{I}\cos\left(\varphi\right)+S_{2}\sqrt{J}\cos\left(\psi\right)\label{eq:Htilde}
\end{equation}
 and
\[
H_{pert}=E_{T}\sqrt{I}\cos\left(\varphi+\left(g_{2}-g_{5}\right)t+\beta\right).
\]

The parameter $\epsilon$ in Eq. (\ref{eq:formal eps}) is used below
to define a hierarchy of Lie transforms, but is set to one at the
end of the calculation. Table (\ref{tab:values BMH}) gives the values
to compute the order of magnitude of $\left|\widetilde{H}\right|$
and $\left|H_{pert}\right|$ respectively. We find $\left|\widetilde{H}\right|\approx3\times10^{-2}$
arcsec/yr, and $\left|H_{pert}\right|\approx9\times10^{-3}$ arcsec/yr.
We perform a canonical change of variables $\left\{ I,J,\varphi,\psi\right\} \rightarrow\left\{ I',J',\varphi',\psi'\right\} $
with Lie transform methods to integrate the term $H_{pert}$ and all
non-resonant harmonics. The procedure is described with all details
in many references \cite{morbidelli1997role,morbidelli2002modern},
but we explain briefly below the general principle.\\

The canonical transformation is given by a function $\chi(I,J,\varphi,\psi)$
such that the new Hamiltonian $H'$ can be computed by
\begin{align*}
H' & =e^{\chi}H,\\
 & =H+\left\{ \chi,H\right\} +\frac{1}{2}\left\{ \chi\left\{ \chi,H\right\} \right\} +...
\end{align*}
where the symbol $\left\{ .\right\} $ represents the canonical Poisson
brackets. The aim is then to choose carefully $\chi$ to eliminate
all non-resonant terms in $H'$. This can be achieved order by order
in $\epsilon$. We expand the function $\chi$ in power of $\epsilon$
as
\[
\chi=\epsilon\chi_{1}+\epsilon^{2}\chi_{2}+...
\]
and we solve order by order in $\epsilon$ the homologic equation
for $\chi_{n}$
\[
\left\{ \chi_{n},H_{int}\right\} +R_{n}=0,
\]
where $H_{int}$ is given by Eq. (\ref{eq:Hint}) and $R_{n}$ gathers
all non-resonant terms of order $\epsilon^{n}$ that are created by
the Lie transforms up to order $n-1$. The procedure leads to the
so-called \emph{resonant normal form}. The Hamiltonian in resonant
normal form only contains terms that can not be integrated out because
they are resonant in the accessible domain of phase space. The resonant
combination of angles up to third order are displayed in Fig (\ref{fig:resonance_map}).\\

\begin{figure}
\begin{centering}
\includegraphics[width=15cm]{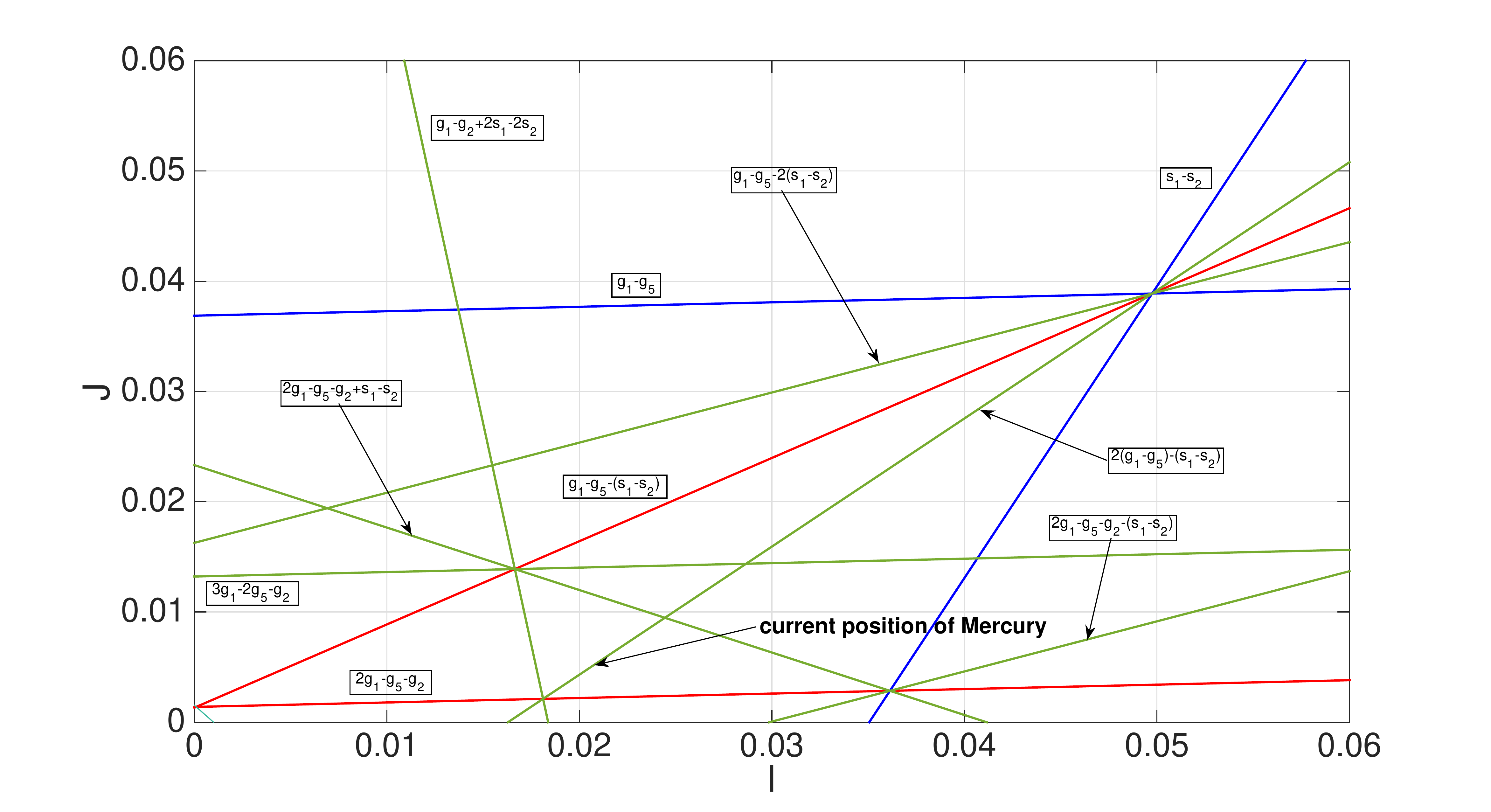}
\par\end{centering}
\caption{The resonance map in action space. The blue lines represent the first
order resonances, the red lines to the second order resonances, and
the green lines to third order. Mercury's current position is close
to an intersection of many third order resonances. \label{fig:resonance_map}}
\end{figure}
The computations of the Lie transforms up to order 3 in $\epsilon$
can be done with the special software TRIP. At each order in $\epsilon$
in the Lie transforms, we keep all terms that involve a resonant angle
in the accessible domain of phase space. The algorithm gives the Hamiltonian
(\ref{eq:BMH}) in terms of the new canonical variables 
\begin{equation}
H'\left(I',J',\varphi',\psi',t\right)=\widetilde{H}'\left(I',J',\varphi',\psi'\right)+\epsilon^{3}H'_{pert}\left(I',J',\varphi',\psi',gt\right)+O(\epsilon^{4}),\label{eq:hamiltonian transform_SI}
\end{equation}
where $\widetilde{H}'$ is the autonomous part of the Hamiltonian,
and $H'_{pert}$ is the part of the Hamiltonian with all resonant
angles of second and third order. The part $H'_{pert}$ has the form
\begin{eqnarray}
H'_{pert}\left(I',J',\varphi',\psi',gt\right) & = & F_{\{2,0,1\}}\left(I',J'\right)\cos\left(2\varphi'+gt\right)\nonumber \\
 & + & F_{\{2,1,1\}}\left(I',J'\right)\cos\left(2\varphi'+\psi'+gt\right)\nonumber \\
 & + & F_{\{1,2,1\}}\left(I',J'\right)\cos\left(\varphi'+2\psi'+gt\right).\label{eq:explicit Hres}
\end{eqnarray}
We have  explicitly computed the coefficients $F_{\{2,0,1\}},F_{\{2,1,1\}},F_{\{1,2,1\}}$
with TRIP, their explicit expression, together with the expression
of $\widetilde{H}'$ are available on request to the authors. 

\subsection{Explicit expression for $D$}

In the present section, we apply stochastic averaging to the dynamics
\begin{align}
\dot{\widetilde{H}'} & =\left\{ H,\widetilde{H}'\right\} \nonumber \\
 & =\epsilon^{3}\left\{ H'_{pert},\widetilde{H}'\right\} .\label{eq:motion good slow_SI}
\end{align}
 to find an order of magnitude for the diffusion of $\widetilde{H}'$.
To simplify the computations and get an explicit expression for the
diffusion coefficient, we have chosen reasonable assumptions. 

We first notice that the terms of largest amplitude in $\widetilde{H}'$
are the terms that depend only on the action variables. To leading
order, the expression of $\widetilde{H}'$ reduces to
\[
\widetilde{H}'(I',J',\varphi',\psi')\approx H_{int}\left(I',J'\right),
\]
with the expression of $H_{int}$ given by Eq. (\ref{eq:Hint}).

Using the above approximation in the right-hand side of (\ref{eq:motion good slow_SI}),
the dynamics of $\widetilde{H}'$ reduces to
\begin{equation}
\dot{\widetilde{H}'}=-\epsilon^{3}\frac{\partial H_{int}}{\partial I'}\frac{\partial H'_{pert}}{\partial\varphi'}-\epsilon^{3}\frac{\partial H_{int}}{\partial J'}\frac{\partial H'_{pert}}{\partial\psi'}.\label{eq:diff coeff1}
\end{equation}
With the expression (\ref{eq:explicit Hres}), Eq. (\ref{eq:diff coeff1})
can be rewritten as

\begin{eqnarray}
\dot{\widetilde{H}'}\left(I',J',\varphi',\psi',gt\right) & = & \overline{F}_{\{2,0,1\}}\left(I',J'\right)\sin\left(2\varphi'+gt\right)\nonumber \\
 & + & \overline{F}_{\{2,1,1\}}\left(I',J'\right)\sin\left(2\varphi'+\psi'+gt\right)\nonumber \\
 & + & \overline{F}_{\{1,2,1\}}\left(I',J'\right)\sin\left(\varphi'+2\psi'+gt\right),\label{eq:explicit coeff slow diffusion}
\end{eqnarray}
where $\left\{ \overline{F}_{\{2,0,1\}},\overline{F}_{\{2,1,1\}},\overline{F}_{\{1,2,1\}}\right\} $
are new coefficients obtained from the expression of $\left\{ F_{\{2,0,1\}},F_{\{2,1,1\}},F_{\{1,2,1\}}\right\} $.
Using stochastic averaging for Eq. (\ref{eq:explicit coeff slow diffusion})
(see e.g. \cite{gardiner1985stochastic}), the long-term evolution
of $\widetilde{H}'$ is equivalent in law to a diffusion process 
\begin{equation}
\dot{\widetilde{H}'}=a\left(\widetilde{H}'\right)+\sqrt{D\left(\widetilde{H}'\right)}\xi(t).\label{eq:diff for slow H}
\end{equation}
The drift term $a\left(\widetilde{H}'\right)$ comes from averaging
Eq. (\ref{eq:explicit coeff slow diffusion}) over fast motion, and
from the correlations between fast and slow motion. Numerical simulations
done with the dynamics (\ref{eq:explicit coeff slow diffusion}) show
that the drift is very small compared to the diffusion, and can be
neglected, at least in the range of timescale of one billion years
we are interested in. In the following, we focus on the diffusion
coefficient $D\left(\widetilde{H}'\right)$. 

The diffusion coefficient can be expressed with a Green-Kubo formula
involving the correlation function of the right-hand side of Eq. (\ref{eq:explicit coeff slow diffusion}).
The complete expression is quite long. In this section, in order to
get reasonable orders of magnitude, we assume that the cross correlations
between different resonant angles give no appreciable contributions.
For example, we neglect correlations such as
\[
\left\langle \overline{F}_{\{2,0,1\}}\left(I'(t),J'(t)\right)\sin\left(2\varphi'(t)+gt\right)\overline{F}_{\{2,1,1\}}\left(I'(0),J'(0)\right)\sin\left(2\varphi'(0)+\psi'(0)\right)\right\rangle .
\]

The functions $\overline{F}(I'(t),J'(t))$ in Eq. (\ref{eq:explicit coeff slow diffusion})
can be decomposed between an non-zero averaged part, and a small perturbation
with zero average. Clearly, the leading order can be computed retaining
only the averaged component of $\overline{F}$. We thus do not longer
take into account the dependance on action variables in (\ref{eq:explicit coeff slow diffusion})
and we systematically replace the functions $\overline{F}\left(I'(t),J'(t)\right)$
by a constant corresponding to their order of magnitude. With the
approximations discussed above, the order of magnitude for $D\left(\widetilde{H}'\right)$
is 
\begin{align}
D\left(\widetilde{H}'\right) & \approx2\left|\overline{F}_{\{2,0,1\}}\right|^{2}\int_{0}^{+\infty}{\rm dt}\left\langle \sin(2\varphi(t)+gt)\sin(2\varphi(0))\right\rangle _{\widetilde{H}'}\nonumber \\
 & +2\left|\overline{F}_{\{2,1,1\}}\right|^{2}\int_{0}^{+\infty}{\rm dt}\left\langle \sin(2\varphi(t)+\psi(t)+gt)\sin(2\varphi(0)+\psi(0))\right\rangle _{\widetilde{H}'}\nonumber \\
 & +2\left|\overline{F}_{\{1,2,1\}}\right|^{2}\int_{0}^{+\infty}{\rm dt}\left\langle \sin(\varphi(t)+2\psi(t)+gt)\sin(\varphi(0)+2\psi(0))\right\rangle _{\widetilde{H}'}.\label{eq:diff approx1}
\end{align}
In Eq. (\ref{eq:diff approx1}), the notation $\left\langle .\right\rangle _{\widetilde{H}'}$
means that the average should be done with a fixed value $\widetilde{H}'$. 

A last approximation is done to compute the correlation functions
of the sinus terms inside the integrals. The two angles $2\varphi+\psi+gt$
and $\varphi+2\psi+gt$ correspond to the resonances $2g_{1}-g_{5}-g_{2}+s_{1}-s_{2}$
and $g_{1}-g_{2}+2(s_{1}-s_{2})$ respectively, and are resonant right
at the center of the accessible domain as displayed in Fig. (\ref{fig:resonance_map}).
Their average frequency is close to zero. On the contrary, the angle
$2\varphi+gt$ is only resonant at the domain boundaries. We choose
to keep only the contribution from the last two terms in the right-hand
side of Eq. (\ref{eq:diff approx1}). Let $\tau_{L}$ be the correlation
time of the angle variables, we choose the approximation 
\[
2\varphi(t)+\psi(t)+gt\approx\varphi(t)+2\psi(t)+gt\approx\theta+W\left(\frac{t}{\tau_{L}}\right),
\]
where $W(t)$ is the standard Brownian motion and $\theta$ is a random
variable with uniform probability distribution over $\left[0,2\pi\right]$.
The term $W\left(\frac{t}{\tau_{L}}\right)$ accounts for the fact
that a resonant angle crosses the resonant conditions and switches
its frequency within a time $\approx\tau_{L}$. We mention that the relation between the Lyapunov exponent of a chaotic
Hamiltonian dynamics with one degree of freedom and two resonances
has been precisely studied by \cite{li2014spin}, but the situation
with two degrees of freedom is more subtle and the results cannot
be directly applied here.The expression (\ref{eq:diff approx1}) for the diffusion coefficient
becomes 
\begin{equation}
D\approx2\left(\left|\overline{F}_{\{2,1,1\}}\right|^{2}+\left|\overline{F}_{\{1,2,1\}}\right|^{2}\right)\int_{0}^{+\infty}\mathbb{E}\left[\sin\left(\theta+W\left(\frac{t}{\tau_{L}}\right)\right)\sin(\theta)\right]{\rm d}t.\label{eq:diff approx2}
\end{equation}
The computation of the integral in (\ref{eq:diff approx2}) is straightforward.
The final result is 
\begin{equation}
D\approx2\left(\left|\overline{F}_{\{2,1,1\}}\right|^{2}+\left|\overline{F}_{\{1,2,1\}}\right|^{2}\right)\tau_{L}.\label{eq:exp for D}
\end{equation}

Finally, we have used the numerical value of the Lyapunov time $\tau_{L}\approx1.1$
Myr obtained with numerical simulations, and we have evaluated numerically
the explicit expressions of $\overline{F}_{\{2,1,1\}}$ and $\overline{F}_{\{1,2,1\}}$.
We get the order of magnitude\textbf{
\begin{equation}
D\approx1.15*10^{-5}\:Myr^{-3}\label{eq:theoretical D}
\end{equation}
}

We further show that the order of magnitude (\ref{eq:theoretical D})
can be obtained in a much more heuristic manner. We have proven that
diffusion of the slow variable $\widetilde{H}'$ is due to third order
secular resonances, that come to order $\epsilon^{3}$ in the Hamiltonian
(\ref{eq:hamiltonian transform_SI}). The order of magnitude for $\overline{F}_{\{2,1,1\}}$
and $\overline{F}_{\{1,2,1\}}$ roughly corresponds to $\frac{\left|H_{pert}\right|^{3}}{\left|\widetilde{H}\right|^{3}}\times\left|\widetilde{H}\right|^{2}$
and expression (\ref{eq:exp for D}) can be written
\begin{equation}
D\approx2\frac{\left|H_{pert}\right|^{6}}{\left|\widetilde{H}\right|^{2}}\tau_{L},\label{eq:simple exp for D_SI}
\end{equation}
where $\left|\widetilde{H}\right|$ is the order of magnitude of the
averaged BMH Hamiltonian. Expression (\ref{eq:simple exp for D_SI})
corresponds to Eq. (9) of the main text, and direct evaluation with
$\left|\widetilde{H}\right|=3.10^{-2}$ arcsec/yr, $\left|H_{pert}\right|\approx9\times10^{-3}$
arcsec/yr, and $\tau_{L}\approx1.1$ Myr gives 
\[
D\approx7.2*10^{-7}\;Myr^{-3}.
\]

\subsection{Destabilization criterion for Mercury's orbit}

We explain in the present section how the stability of Mercury's orbit
can be directly related to the value of the slow variable $\widetilde{H}'$.
$\widetilde{H}'$ is obtained by Lie transforms of $\widetilde{H}$
given by (\ref{eq:Htilde}). The explicit expression of $\widetilde{H}'$
is thus composed of a part that depends only on action variables,
and a large number of periodic terms that involve the angle variables.
The leading terms in the action-dependent part of $\widetilde{H}'$
is given by $H_{int}$. We can thus crudely write the decomposition
\[
\widetilde{H}'(I,J,\varphi,\psi,t)=H_{int}\left(I,J\right)+G\left(I,J,\varphi,\psi,t\right),
\]
 where $G$ is some intricate function. For any fixed value of $\widetilde{H}'$,
the variations of $H_{int}$ are bounded between an upper and a lower
value
\[
\Gamma_{inf}\left(\widetilde{H}'\right)\leq H_{int}\leq\Gamma_{sup}\left(\widetilde{H}'\right)
\]
 that depend in a non-trivial way of the maximal amplitude of $G\left(I,J,\varphi,\psi,t\right)$.
The destabilization criterion $H_{int}=H_{cr}$ thus translates into
the equivalent criterion 
\[
\Gamma_{inf}\left(\widetilde{H}'\right)=H_{cr}.
\]
 Let us call $h_{cr}$ the value such that $\Gamma_{inf}\left(h_{cr}\right)=H_{cr}$,
Mercury's destabilization is directly related to the event $\widetilde{H}'=h_{cr}$.
This argument shows why destabilization of Mercury is directly related
to the event $\widetilde{H}'$ hitting the threshold value $h_{cr}$
.

Given the complexity of the explicit expressions of $\widetilde{H}'$
and $\Gamma_{inf}\left(\widetilde{H}'\right)$, the destabilization
criterion has to be treated in an empirical manner. The value of $\widetilde{H}'$
is better replaced by the local time average $h(t)=\left\langle \widetilde{H}\right\rangle _{[t-\theta,t+\theta]}$,
where the time frame of length $\theta$ should satisfy $\theta\gg\frac{1}{g_{2}-g_{5}}$.
This approximation is described precisely in the main text. In practice,
we have chosen $\theta=2$ Myr. To compute the threshold value $h_{cr}$
, we also use a numerical approach: we record the values of $h(t_{cr})$
at the destabilization time, for a large number of destabilized trajectories.
The distribution of $h(t_{cr})$ is peaked at a particular value,
thus confirming the existence of the threshold $h_{cr}$. We find
$h_{cr}\approx-0.048$ arcsec/yr.

\section{Explicit expression for the distribution of first exit times of a
Brownian motion from a bounded domain \label{sec:Resolution-of-the}}

In the present section, we show how to derive the probability distribution
function $\rho(\tau)$ of first exit time of a standard Brownian motion
from the domain $[h_{cr},h_{sup}]$, starting at $h_{0}$ and with
reflective condition at $h=h_{sup}$. 

Let $G\left(h,t\right):=\int_{h_{cr}}^{h_{sup}}\mathbb{P}\left(h',t|h,0\right){\rm d}h'$
be the probability that the Brownian particle starting at $h$ is
still in the domain $\left[h_{cr},h_{sup}\right]$ at time $t$. It
can be shown that the distribution $G(h,t)$ satisfies the same diffusion
equation as $\mathbb{P}\left(h',t|h,0\right)$ (see \cite{gardiner1985stochastic})
\begin{equation}
\frac{\partial G}{\partial t}=D\frac{\partial^{2}G}{\partial h^{2}}.\label{eq:appendix1}
\end{equation}
At time $t=0$, the particle is inside the domain, which means that
$G(h,0)=1$ for all $h\in\left[h_{cr},h_{sup}\right]$. The absorbing
boundary condition at $h=h_{cr}$ and the reflecting boundary condition
at $h=h_{sup}$ can be equivalently expressed with the distribution
$G$ as 
\begin{equation}
\textrm{for all }t>0,\:\begin{cases}
G(h_{cr},t) & =0,\\
\frac{\partial G}{\partial h}\left(h_{sup},t\right) & =0.
\end{cases}\label{eq:appendix2}
\end{equation}
We solve the problem (\ref{eq:appendix1}-\ref{eq:appendix2}) by
decomposing the solution into proper modes. Let us introduce the standard
scalar product 
\[
\left\langle f,g\right\rangle =\frac{2}{h_{sup}-h_{cr}}\int_{h_{cr}}^{h_{sup}}f(h)g(h){\rm d}x.
\]
It can be checked that the family of functions 
\[
e_{n}(h)=\cos\left(\pi\left(n+\frac{1}{2}\right)\frac{h-h_{sup}}{h_{cr}-h_{sup}}\right)\textrm{ with }n\in\mathbb{N}
\]
form an orthonormal basis of all functions $G(x,t)$ satisfying the
boundary conditions (\ref{eq:appendix2}). The solution of (\ref{eq:appendix1}-\ref{eq:appendix2})
can thus be expressed as the Fourier series
\begin{equation}
G(h,t)=\stackrel[n=0]{+\infty}{\sum}g_{n}(t)e_{n}(h),\label{eq:fourier dec}
\end{equation}
where the coefficients $g_{n}(t)$ are defined as the projection of
$G$ on the orthonormal basis, that is $g_{n}(t):=\left\langle G(h,t)e_{n}(h)\right\rangle $.
Using the Fourier decomposition (\ref{eq:fourier dec}), we find that
$G$ is solution of (\ref{eq:appendix1}) if and only if 
\[
g_{n}(t)=g_{n}(0)e^{-\pi^{2}\left(n+\frac{1}{2}\right)^{2}\frac{D}{\left(h_{sup}-h_{cr}\right)^{2}}t}.
\]
The value $g_{n}(0)$ can be found with the initial condition $G(h,0)=1$.
We get
\[
g_{n}(0)=\left\langle G(h,0)e_{n}(h)\right\rangle =\frac{2}{\pi}\frac{(-1)^{n}}{n+\frac{1}{2}}.
\]
Finally, the solution $G(h,t)$ can be expressed explicitly as 
\begin{equation}
G(h,t)=\frac{2}{\pi}\stackrel[n=0]{+\infty}{\sum}\frac{\left(-1\right)^{n}}{n+\frac{1}{2}}\cos\left(\pi\left(n+\frac{1}{2}\right)\frac{h-h_{sup}}{h_{cr}-h_{sup}}\right)e^{-\pi^{2}\left(n+\frac{1}{2}\right)^{2}\frac{D}{\left(h_{sup}-h_{cr}\right)^{2}}t}.\label{eq:cumulproba}
\end{equation}
As $G(h,t)$ is the probability to be still in the domain $[h_{cr},h_{sup}]$
at time $t$, it is related to $\rho(\tau)$ by
\[
G(h,t)=\int_{t}^{+\infty}\rho(\tau){\rm d}\tau.
\]
Therefore, the time derivative of Eq. (\ref{eq:cumulproba}) gives
the explicit expression of $\rho(\tau)$
\begin{equation}
\rho(\tau)=\frac{2\pi D}{\left(h_{sup}-h_{cr}\right)^{2}}\stackrel[n=0]{+\infty}{\sum}\left(-1\right)^{n}\left(n+\frac{1}{2}\right)\cos\left(\pi\left(n+\frac{1}{2}\right)\frac{h_{0}-h_{cr}}{h_{sup}-h_{cr}}\right)\exp\left(-\pi^{2}\left(n+\frac{1}{2}\right)^{2}\frac{D\tau}{\left(h_{sup}-h_{cr}\right)^{2}}\right).\label{eq:exp rhoth}
\end{equation}
This expression is used for the fit in Fig. (4) of the main text.

\section{Average and variance of a Brownian bridge}

In the present section, we show how to obtain explicitly the red curves
in Fig. (5) of the main text.

The aim is to compute the probability
\begin{equation}
\rho^{\tau_{ex}}(h,t):=\mathbb{P}\left(h,t|\left\{ h_{0},0\right\} \cap\left\{ \tau=\tau_{ex}\right\} \right)\label{eq: def proba instanton}
\end{equation}
 to have a trajectory at location $h$ at time $t$ with the constrains
that the trajectory starts at $h_{0}$ and exits the domain at time
$\tau=\tau_{ex}$, for a standard Brownian motion of diffusion coefficient
$D$. The inequality $0<t<\tau_{ex}$ should be satisfied. Using Bayes
theorem and Markov property, the probability distribution (\ref{eq: def proba instanton})
can be written as 
\begin{equation}
\rho^{\tau_{ex}}(h,t)=\frac{\mathbb{P}\left(\tau=\tau_{ex}|h,t\right)\mathbb{P}\left(h,t|h_{0},0\right)}{\mathbb{P}\left(\tau=\tau_{ex}|h_{0},0\right)}.\label{eq:proba instanton decomposed}
\end{equation}
All probability distributions in the right-hand side of (\ref{eq:proba instanton decomposed})
have explicit expressions. The probability $\mathbb{P}\left(\tau=\tau_{ex}|h,t\right)$
to exit the domain starting at a given position can be obtained from
equation (\ref{eq:exp rhoth}) in the limit $h_{sup}\rightarrow+\infty$.
We have thus 
\begin{align*}
\mathbb{P}\left(\tau=\tau_{ex}|h,t\right) & =\frac{1}{\tau_{ex}-t}\frac{h-h_{cr}}{\sqrt{4\pi D\left(\tau_{ex}-t\right)}}e^{-\frac{\left(h-h_{cr}\right)^{2}}{4D\left(\tau_{ex}-t\right)}},\\
\mathbb{P}\left(\tau=\tau_{ex}|h_{0},0\right) & =\frac{1}{\tau_{ex}}\frac{h_{0}-h_{cr}}{\sqrt{4\pi D\tau_{ex}}}e^{-\frac{\left(h_{0}-h_{cr}\right)^{2}}{4D\tau_{ex}}}.
\end{align*}
The last term $\mathbb{P}\left(h,t|h_{0},0\right)$ is simply the
solution of the free diffusion equation in an infinite domain, which
is the classical result
\[
\mathbb{P}\left(h,t|h_{0},0\right)=\frac{1}{\sqrt{4\pi Dt}}e^{-\frac{\left(h-h_{0}\right)^{2}}{4Dt}}.
\]
After some algebra, we obtain the following explicit expression for
$\rho^{\tau_{ex}}(h,t)$ (valid for $h>h_{cr}$ and $0<t<\tau_{ex}$)
\begin{align}
\rho^{\tau_{ex}}(h,t) & =\frac{h-h_{cr}}{h_{0}-h_{cr}}\left(\frac{\tau_{ex}}{\tau_{ex}-t}\right)^{3/2}\frac{1}{\sqrt{4\pi Dt}}\label{eq:explicit instanton distribution}\\
 & \times\left\{ \exp\left(-\frac{\left(h-h_{cr}-(1-s)\left(h_{0}-h_{cr}\right)\right)^{2}}{4D\tau_{ex}s(1-s)}\right)-\exp\left(-\frac{\left(h-h_{cr}+(1-s)\left(h_{0}-h_{cr}\right)\right)^{2}}{4D\tau_{ex}s(1-s)}\right)\right\} ,\nonumber 
\end{align}
where we have introduced the ratio $s=\frac{t}{\tau_{ex}}$. It can
be quite easily checked that $\int_{h_{cr}}^{+\infty}\rho^{\tau_{ex}}(h,t){\rm d}h=1$,
because $\rho^{\tau_{ex}}(h,t)$ is a probability density. 

The instanton trajectory, and the variance of the distribution around
the instanton can be obtained with the first and the second moments
of the distribution (\ref{eq:explicit instanton distribution}). We
define the average trajectory $\left\{ \bar{h}(t)\right\} _{0<t<\tau_{ex}}$
as 
\begin{equation}
\bar{h}(t)=\int_{h_{cr}}^{+\infty}h\rho^{\tau_{ex}}(h,t){\rm d}h.\label{eq:instanton}
\end{equation}
There is a small difference between the average trajectory defined
by (\ref{eq:instanton}) and the instanton trajectory $\widetilde{h}(t)$
which is the trajectory of highest probability. The trajectory of
highest probability is the straight trajectory of equation
\[
\widetilde{h}(t)=\frac{t}{\tau_{ex}}h_{cr}+\left(\frac{\tau_{ex}-t}{\tau_{ex}}\right)h_{0}.
\]

The distribution of trajectories that exit the domain for short times
is more and more concentrated around the trajectory of highest probability
when $\tau_{ex}$ goes to zero. To first approximation, $\widetilde{h}\approx\bar{h}$
when $\tau_{ex}$ is small compared to $\tau^{*}$. However, the average
trajectory is a bit curved when $t$ gets closer to $\tau_{ex}$ because
of the influence of the absorbing boundary condition. We represent
in Fig. (5) of the main text the averaged trajectory instead of the
instanton trajectory because it can more easily be compared to numerical
results. To study the trajectories dispersion around the instanton,
we can also compute the standard deviation 
\begin{equation}
\delta\bar{h}(t)=\left[\int_{h_{cr}}^{+\infty}\left(h-\bar{h}(t)\right)^{2}\rho^{\tau_{ex}}(h,t){\rm d}h\right]^{1/2}.\label{eq:variance instanton}
\end{equation}

Expressions (\ref{eq:instanton}) and (\ref{eq:variance instanton})
can be evaluated numerically. The three red curves in Fig. (5) of
the main text are those of equations (from highest to lowest) $h(t)=\bar{h}(t)+\delta\bar{h}(t)$,
$h(t)=\bar{h}(t)$ and $h(t)=\bar{h}(t)-\delta\bar{h}(t)$ respectively,
for $\tau_{ex}=445$ million years.\bibliographystyle{plain}

\section{Information about the numerical simulations}
\subsection{Figure 4}
The probability distribution of Mercury's first destabilization time represented by the blue curve in Fig. 4 of the main text has been obtained from a direct numerical simulation of Hamilton's equations 
\begin{equation}
\begin{cases}
\dot{\varphi}(t) & =\frac{\partial H}{\partial I}\\
\dot{\psi}(t) & =\frac{\partial H}{\partial J},
\end{cases}
\end{equation}
where $H$ is given by Eq. (\ref{eq:BMH}). We used a Runge-Kutta scheme of order 4. We integrated $N=126518$ trajectories with initial conditions uniformly chosen in the range $[\varphi_0-10^{-2},\varphi_0+10^{-2}]$.  The simulation is stopped either when the trajectory reaches $I=0.06$ or when the time of integration becomes larger than $T=3.1$ billion years. We recorded $n=57330$ trajectories that have reached  $I=0.06$ before the maximal integration time. We recall that $I=0.06$ means that the trajectory has entered the unbounded part of phase space and the orbit can therefore be considered as destabilized.

Then, the blue curve of Fig. (4) of the main text  is obtained by fitting the distribution of the $n$ recorded times with the function "kernel" of matlab, with the normalization set to one. To compare the result with the diffusion model, we plotted in red in Fig (4) of the main text the expression $\rho(\tau)*N/n$ with $\rho(\tau)$ given by Eq.(\ref{eq:exp rhoth}) and $D=9.6*10^{-7}$ $Myr^{-3}$.

\subsection{Estimation of the drift coefficient}
In this section, we give an estimation of the drift $a\left(\widetilde{H}'\right)$ in Eq. (\ref{eq:diff for slow H}). The aim is to show that $a\left(\widetilde{H}'\right)$ can be neglected to compute Mercury's first destabilization time with the diffusion model. For this purpose, we do the following numerical simulation: we integrate Hamilton's equations for the dynamics defined by the Hamiltonian of Eq. (4) of the main text
\[
\widetilde{H}=H_{int}+E_2\sqrt{I}\cos\left(\varphi\right)+S_2\sqrt{J}\cos\left(\psi\right).
\]
We record $N=1000$ trajectories starting from initial conditions chosen uniformly in the range $[\varphi_0-10^{-2},\varphi_0+10^{-2}]$, and for a time $T=206$ Myr. Note that those trajectories are necessarily bounded because the value of $\widetilde{H}$ is conserved. Then we compute the explicit expression of $\dot{\widetilde{H}}(I,J,\varphi ,\psi ,t)=\left\{\widetilde{H},H_{pert}\right\}$, and we use it to integrate the equation
\[
\dot{F}=\dot{\widetilde{H}}(I(t),J(t),\varphi(t) ,\psi(t) ,t)\label{eq:adiabatic},
\]
where $(I(t),J(t),\varphi(t) ,\psi(t))$ is a trajectory computed previously. $F(t)$ can be seen as a good approximation for $\widetilde{H}' (t)-\widetilde{H}' (0)$ at short times. We obtain this way a set of $N=1000$ trajectories $\{F_i(t)\}$.  

Finally, we plot on Fig. (\ref{fig:adiabatic}) the histogram of the $\{F_i(t)\}$ for $t=41/82/123/165/206$ Myr, and we fit the different histograms with the Gaussian distribution
\[
\rho(F)=\frac{1}{\sqrt{2\pi \sigma^2(t)}}e^{-\frac{\left(F-m(t)\right)^{2}}{2\sigma^2(t)}}.
\]
We observe that the quantity $\sigma^2(t)$ indeed scales linearly with $t$, as expected for a diffusion process.The value of $m(t)$ is non vanishing because of the short-term oscillations of $\widetilde{H}$ on the Myr timescale. To obtain a relevant order of magnitude for the long-term drift, we have to subtract the shift due to the short-term oscillations. The quantity
\[
A=\max\{m(t)-m(t');t,t'\in[0,T]\}/T \approx 1.1*10^{-6}\;Myr^{-2}
\]
gives us an order of magnitude for the drift $a\left(\widetilde{H}'\right)$ in Eq. (\ref{eq:diff for slow H}). Correspondingly, we find that the error due to the drift, for $t=3000\;Myr $ should not exceed $At\approx 3.3*10^{-3}\; Myr^{-1}$. With $D\approx 10^{-6}\;Myr^{-3}$, we find that the variation of $\widetilde{H}' (t)$ due to the diffusion coefficient over the same timescale  is of the order of $\sqrt{Dt}\approx 5*10^{-2}\; Myr^{-1}$. We conclude that there is one order of magnitude between the respective effects of the drift term and the diffusion term in Eq. (\ref{eq:diff for slow H}), and that the former can be neglected on the billion years timescale.
\begin{figure}
\begin{centering}
\includegraphics[width=15cm]{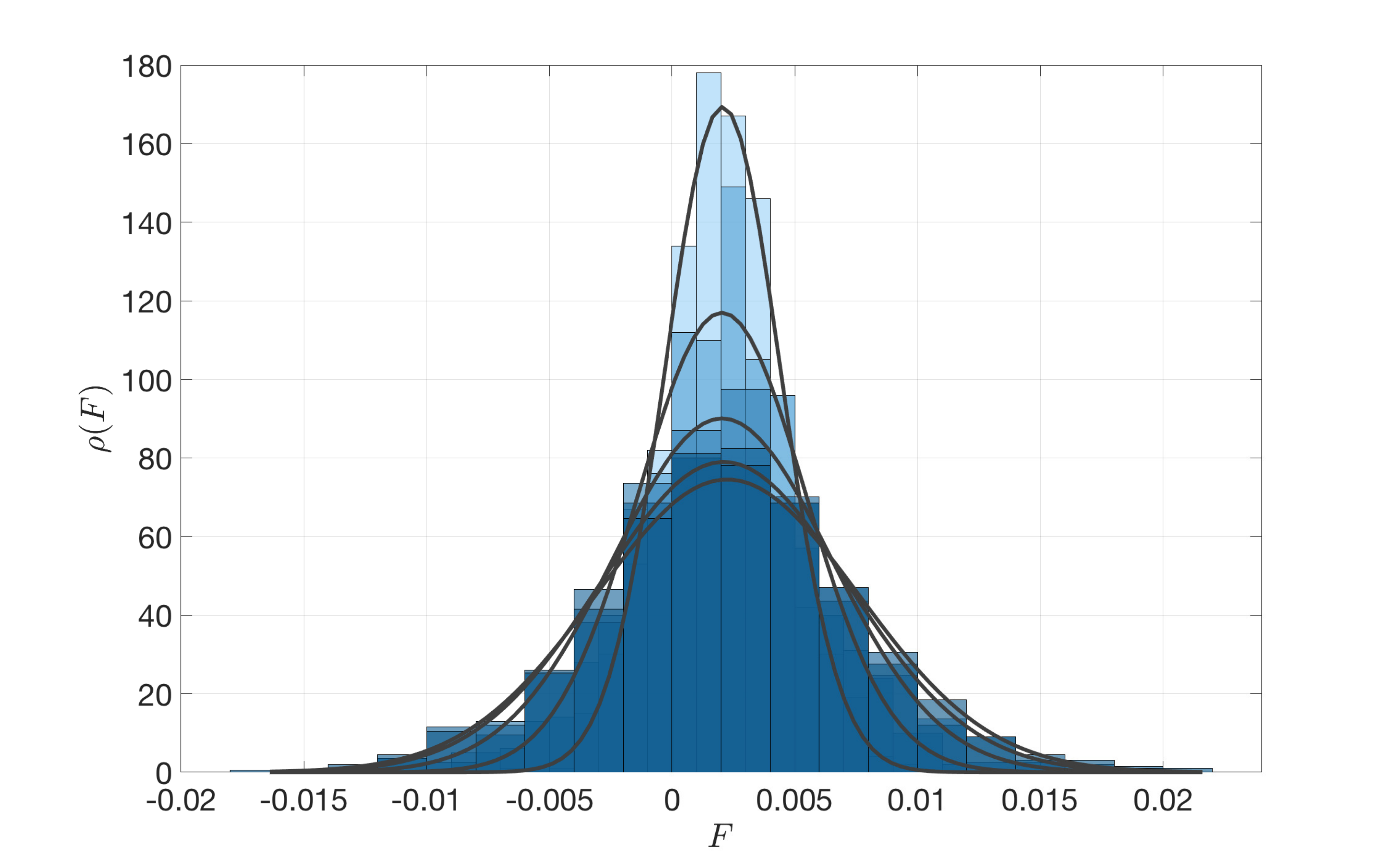}
\par\end{centering}
\caption{The distribution of $F(t)$ for different times $t=41/82/123/165/206$ Myr.  The blue histograms represent the distributions of the $N=1000$ trajectories generated by Eq. (\ref{eq:adiabatic}). The darkness of the histogram increases with time. The grey curves represent the Gaussian fits of the histograms. The average and variance of those distributions give us order of magnitudes for the drift  and the diffusion coefficient of the diffusion model for $\widetilde{H}'$. To leading order, the drift can be neglected.  \label{fig:adiabatic}}
\end{figure}

\end{widetext}

\end{document}